\documentclass[twocolumn, amsmath, superscriptaddress]{revtex4}
\usepackage{graphicx}
\usepackage{amsmath}
\usepackage{amssymb}
\usepackage{xcolor}
\usepackage{bm}

\begin{document}

\title{Random matrix approach to plasmon resonances in the random impedance network model of disordered nanocomposites}

\author{N.\,A.\,Olekhno}
\affiliation{ITMO University, 49 Kronverksky Pr. St. Petersburg, 197101 Russia}
\email{olekhnon@gmail.com}
\author{Y.\,M.\,Beltukov}
\affiliation{Ioffe Institute, Politekhnicheskaya ul. 26, St. Petersburg, 194021 Russia}

\date{\today}

\begin{abstract}
Random impedance networks are widely used as a model to describe plasmon resonances in disordered metal-dielectric and other two-component nanocomposites. In the present work, the spectral properties of resonances in random networks are studied within the framework of the random matrix theory. We have shown that the appropriate ensemble of random matrices for the considered problem is the Jacobi ensemble (the MANOVA ensemble). The obtained analytical expressions for the density of states in such resonant networks show a good agreement with the results of numerical simulations in a wide range of metal filling fractions $0<p<1$. A correspondence with the effective medium approximation is observed.
\end{abstract}

\maketitle

\begin{figure}[t]
    \includegraphics[width=8cm]{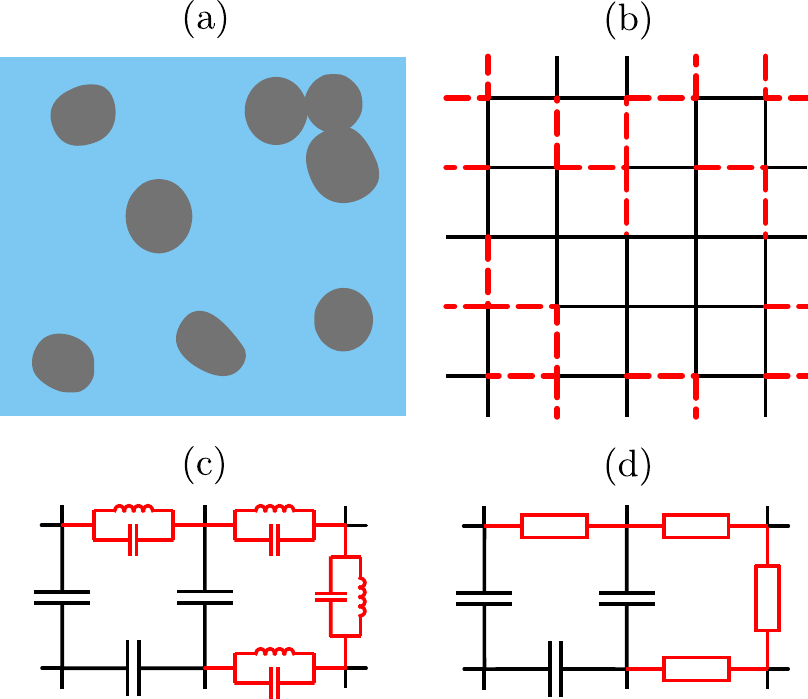}
    \caption{(Color online) (a) Sketch of a disordered nanoparticle composite. (b) A random impedance network. ``Metallic'' bonds are shown with red (gray) lines, and ``dielectric'' bonds are shown with black lines. (c), (d) Examples of particular impedance network models.}
    \label{fig:network}
\end{figure}

\emph{Introduction.} Disordered metal-dielectric nanocomposites form a type of optical metamaterials which are relatively simple to fabricate [Fig.~\ref{fig:network}(a)]. Their geometries vary from a dielectric medium with metallic inclusions of a submicron size to a metallic medium with dielectric holes, depending on the metal fraction. Such systems demonstrate a lot of interesting optical phenomena assisted by surface plasmon resonances in metallic regions, namely, surface-enhanced Raman scattering (SERS) \cite{Le_Ru_2009}, high-harmonic generation \cite{2001_Breit}, and the Purcell effect \cite{2015_Carminati}. Various nonlinearities in such composites especially increase near the percolation threshold \cite{2000_Sarychev}.

A number of classical models of the percolation theory are based on random impedance networks [Fig.~\ref{fig:network}(b)] which have been widely applied to study the transport properties \cite{1973_Kirkpatrick, 1990_Clerc} and resonances \cite{1992_Bergman, 2000_Sarychev} in disordered nanocomposites. Fluctuations of the local electric fields responsible for SERS  have been considered in the framework of the random impedance network model \cite{2000_Sarychev, 1997_Brouers, 1998_Brouers}, as well as the density of states (DOS) \cite{1998_Jonckheere, 2000_Albinet, 2012_Murphy} and the optical absorption \cite{1987_Koss, 1993_Zhang, 1995_Zhang}. Such models demonstrate the presence of the Anderson transition \cite{1999_Sarychev, 2017_Murphy} and possess multifractal properties of electric field distributions \cite{1998_Jonckheere, 2002_Gu}. However, the main part of the results is obtained numerically.

In the present Rapid Communication, we propose to apply the random matrix theory for a unified description of the DOS in random impedance networks, which are widely used as a model of plasmon resonances in disordered nanocomposites \cite{2000_Sarychev}. The random matrix theory has found numerous applications in different branches of physics, for example, in nuclear physics \cite{1962_Dyson}, quantum chaos \cite{Haake}, description of the conductance of disordered channels \cite{Oxford_Handbook, 2000_Mirlin}, coherent perfect absorbers \cite{2017_Li}, and the mechanical properties of disordered solids \cite{2015_Beltukov}. The random matrix theory was also applied to study the statistical properties of financial markets and computer networks \cite{Oxford_Handbook}. Each of the mentioned problems has some important symmetries, which lead to different symmetry classes of random matrices (the so-called \emph{random matrix ensembles}) \cite{2008_Evers}.

\emph{Random impedance network model.} We consider a widely used quasistatic approach when the electric field is assumed to be curl-free ($\operatorname{rot}{\mathbf E}={\mathbf 0}$), and the electrostatic potential $\varphi$ can be introduced such that ${\mathbf E} = -\operatorname{grad}\varphi$. Since a characteristic size of inclusions is about tens of nanometers, this condition can be satisfied in THz, infrared, and even visible optical regions. It is well known that Maxwell's equations are reduced to the equation for an eddy current $\operatorname{div}\mathbf{j}=0$ within the quasistatic approach \cite{1973_Kirkpatrick, 1977_Webman}. For a given frequency $\omega$, the current $\mathbf{j}$ and the electric field $\mathbf{E}$ are related by the material equation $\mathbf{j}(\omega, \mathbf{r})=\sigma(\omega, \mathbf{r})\mathbf{E}(\omega, \mathbf{r})$. The conductivity $\sigma(\omega, \mathbf{r})$ is related to the permittivity $\varepsilon(\omega, \mathbf{r})$ of the same region as $\sigma(\omega, \mathbf{r})=i\omega\varepsilon(\omega, \mathbf{r})/4\pi$ \cite{LL8}. The obtained equations are discretized on a mesh. For that reason, square \cite{2000_Sarychev, 1996_Clerc, 1998_Jonckheere, 1977_Webman} and simple cubic \cite{2000_Albinet} lattices have been used. After a discretization, the equations $\operatorname{div}\bm{j}=0$ and $\operatorname{rot}\bm{E}={\mathbf 0}$  transform to the first and the second Kirchhoff's rules, respectively. One can simultaneously represent both Kirchhoff's rules as a linear system,
\begin{equation}
    \sum_{j=1}^{N} g_{ij}(\omega)\varphi_{j} = 0,   \label{eq:Kirchhoff}
\end{equation}
with $\varphi_{j}$ being the electric potential at site $j$ and $g_{ij}(\omega)$ being a complex conductance of the bond between sites $i$ and $j$ in a network with $N$ sites \cite{1973_Kirkpatrick}. Diagonal entries $g_{ii}(\omega)$ are defined as $g_{ii}(\omega) = -\sum_{i \neq j}g_{ij}(\omega)$.

Next, we briefly consider some of the simplest models. In the optical frequency range, the permittivity of a metal can be described with the aid of the Drude model $\varepsilon_m(\omega)=1-\omega_p^2/\omega^2$, with $\omega_p$ being a plasma frequency of a metal. At the same time, the permittivity of dielectric regions can be taken as a constant $\varepsilon_d$. Then, a composite is replaced with a resonant $LC$ network [Fig.~\ref{fig:network}(c)] \cite{1989_Zeng, 1995_Zhang, 2016_OBP}. At low frequencies another model having the form of an $RC$ network can be introduced [Fig.~\ref{fig:network}(d)]. Such a model is used to consider the transient responses in composites \cite{1990_Clerc, 2009_Korniss, 2017_Aouaichia}.

In order to study the properties of the resonances, we reduce Maxwell's equations to the eigenvalue problem \cite{1979_Bergman_1, 1979_Bergman_2}. In a generic two-component system, the conductance $g_{ij}$ can be represented as 
\begin{equation}
    g_{ij}(\omega) = \sigma_m(\omega)M_{ij} + \sigma_d(\omega)D_{ij}.  \label{eq:2c}
\end{equation}
Matrices $M$ and $D$ are defined in the following manner. In the general case, we can assume that $M_{ij} = -1$ if sites $i$ and $j$ are connected by a metallic bond and $D_{ij}=-1$ if sites $i$ and $j$ are connected by a dielectric bond. The remaining off-diagonal elements of matrices $M$ and $D$ are zero. Diagonal elements are defined as $M_{ii} = -\sum_{j\neq i} M_{ij}$ and $D_{ii} = -\sum_{j\neq i} D_{ij}$. Thus, matrices $M$ and $D$ represent discrete Laplacians defined on corresponding metallic and dielectric subsets.

For certain frequencies $\omega=\omega_j$, the linear system (\ref{eq:Kirchhoff}) has nontrivial solutions $\varphi_j$, which represent dielectric resonances in the network \cite{1992_Bergman} corresponding to plasmon resonances of a composite. For a two-component system (\ref{eq:2c}), eigenfrequencies can be found using the generalized eigenvalue problem \cite{1998_Jonckheere}
\begin{equation}
    M\varphi_j = \lambda_j (M + D)\varphi_j,   \label{eq:GEP}
\end{equation}
where eigenvalues $\lambda_j$ are related to eigenfrequencies $\omega_j$ as
\begin{equation}
    \lambda_j = \frac{\sigma_d(\omega_j)}{\sigma_d(\omega_j) - \sigma_m(\omega_j)} = \frac{\varepsilon_d(\omega_j)}{\varepsilon_d(\omega_j) - \varepsilon_m(\omega_j)}.   \label{eq:lambda}
\end{equation}
Eigenvalues $\lambda_j$ and eigenvectors $\varphi_j$ are determined by matrices $M$ and $D$, which do not depend on the dielectric functions of constituents $\varepsilon_{m,d}(\omega)$. As a result, the dielectric functions $\varepsilon_{m,d}(\omega)$ affect only eigenfrequencies $\omega_j$, which are related to eigenvalues $\lambda_j$ by Eq.~(\ref{eq:lambda}). It is important to mention that matrices $M$ and $D$ are positive semidefinite \cite{Horn-book}, and thus $0 \le \lambda_j \le 1$ for networks of an arbitrary geometry \cite{Ortega-book}.

Considerable efforts have been put forth to figure out an analytic description of the DOS $\rho(\lambda) = \frac{1}{N}\sum_{j=1}^{N}{\delta(\lambda-\lambda_j)}$ in such networks within the framework of the random matrix theory (RMT) \cite{1999_Fyodorov, 1999_Fyodorov_1, 2001_Fyodorov, 2003_Staring}. In the above-mentioned papers, Gaussian ensembles of random matrices have been applied to describe resonances in long-range networks with a quasi-one-dimensional topology, which show no direct relation to the problem of plasmon resonances in two-dimensional and three-dimensional disordered nanocomposites.

In order to simplify the problem, we will consider a common model of a random network, which assumes that each bond in a lattice with a coordination number $z$ is {\it metallic} with probability $p$ or {\it dielectric} with probability $1-p$ \cite{1998_Jonckheere}.

\emph{Density of states.} The matrices $M$ and $D$ are positive semidefinite, so they can be represented in the form $M = AA^T$ and $D = BB^T$. There are different possibilities to choose matrices $A$ and $B$ for the same matrices $M$ and $D$. However, there is the most natural form of matrices $A$ and $B$, which is known as the incidence matrix in the graph theory~\cite{Bollobaas}. In this case, the height of matrix $A$ is the number of sites and the width of matrix $A$ is the number of metallic bonds in the lattice. The nonzero matrix elements are $A_{ki}=1$ and $A_{kj}=-1$, where $k$ is the index of a bond and $i$ and $j$ are indices of sites connected by the $k$th bond. For each pair of $i$ and $j$, the choice of $1$ and $-1$ is arbitrary. The definition of matrix $B$ is the same but for dielectric bonds. Therefore, the generalized eigenvalue problem (\ref{eq:GEP}) can be written in the form
\begin{equation}
    |AA^T - \lambda(AA^T + BB^T)|=0.   \label{eq:AB}
\end{equation}
In the above definition, matrices $A$ and $B$ are sparse with a certain structure of nonzero elements. However, it does not play a crucial role for the DOS. Indeed, for any orthogonal matrices $U$, $V$, and $W$, we can introduce matrices $\widetilde{A} = UAV$ and $\widetilde{B} = UBW$, which leads to the generalized eigenvalue problem
\begin{equation}
    |\widetilde{A}\widetilde{A}^T - \lambda (\widetilde{A}\widetilde{A}^T + \widetilde{B}\widetilde{B}^T)| = 0,   \label{eq:Jacobi}
\end{equation}
with the same set of eigenvalues $\lambda_j$ as for Eq.~(\ref{eq:AB}). As a result, one can assume that the DOS mostly depends on the correlations given by the form of Eq.~(\ref{eq:Jacobi}) rather than the internal correlations of matrices $A$ and $B$ \cite{2015_Beltukov}. Thus, we assume that matrices $\widetilde{A}$ and $\widetilde{B}$ are Gaussian random matrices. The sizes of the matrices are $N\times K_m$ and $N\times K_d$, respectively, where $K_m = pzN/2$ and $K_d = (1-p)zN/2$ are the total numbers of metallic and dielectric bonds, and $N$ is a number of sites in the lattice. In this case, Eq.~(\ref{eq:Jacobi}) defines the so-called Jacobi ensemble of the random matrix theory \cite{Forrester}. It is also known as the MANOVA ensemble since Eq. (\ref{eq:Jacobi}) has a special meaning in the multivariate analysis of variance (MANOVA).

For the Jacobi ensemble, the joint probability distribution of an ascending list of eigenvalues $\lambda_j$ is
\begin{multline}
    p(\lambda_1, \ldots, \lambda_N) = C\prod_i \lambda_i^{(K_m - N - 1)/2}\\
    \times \prod_i(1-\lambda_i)^{(K_d - N - 1)/2}\prod_{i<j} (\lambda_j - \lambda_i), \label{eq:joint}
\end{multline}
where $C$ is a normalization constant \cite{1963_Constantine}. The last product in Eq.~(\ref{eq:joint}) vanishes when $\lambda_i=\lambda_j$. This leads to the level repulsion effect which is well known for the Gaussian orthogonal ensemble (GOE) and was also observed for random impedance networks~\cite{2017_Murphy}. However, eigenvalue probability density functions (i.e., DOS) for the Jacobi ensemble and for the GOE are different. For the Jacobi ensemble, it has the form \cite{Forrester, 2013_Erdos}
\begin{equation}
    \rho(\lambda) = z\frac{\sqrt{(\lambda-\lambda_{-})(\lambda_{+}-\lambda)}}{4\pi \lambda (1-\lambda)}, \quad \lambda_- \leq \lambda \leq \lambda_+   \label{eq:RMT_DOS}
\end{equation}
where the spectral edges $\lambda_{\pm}$ are given by
\begin{equation}
    \lambda_\pm = p + \frac{2-4p}{z} \pm \frac{2}{z}\sqrt{2p(1-p)(z-2)}.
\end{equation}
In addition to eigenvalues defined by $\rho(\lambda)$, there is a number of degenerate eigenvalues $\lambda=0$ and $\lambda=1$. The relative number of eigenvalues $\lambda=0$ is $n_0 = 1 - K_m/N = 1 - zp/2$, and the relative number of eigenvalues $\lambda=1$ is $n_1 = 1 - K_d/N = 1 - z(1-p)/2$.

One of the fundamental properties of the generalized eigenvalue problem (\ref{eq:GEP}) is the {\it homogeneity symmetry} \cite{1998_Jonckheere}: The DOS obeys the relation $\rho(p, \lambda) = \rho(1-p, 1-\lambda)$ due to the equivalence of the statistical properties of matrices $M$ and $D$. It is obvious that the Jacobi ensemble satisfies this symmetry.

\begin{figure}[t]
    \includegraphics[width=8cm]{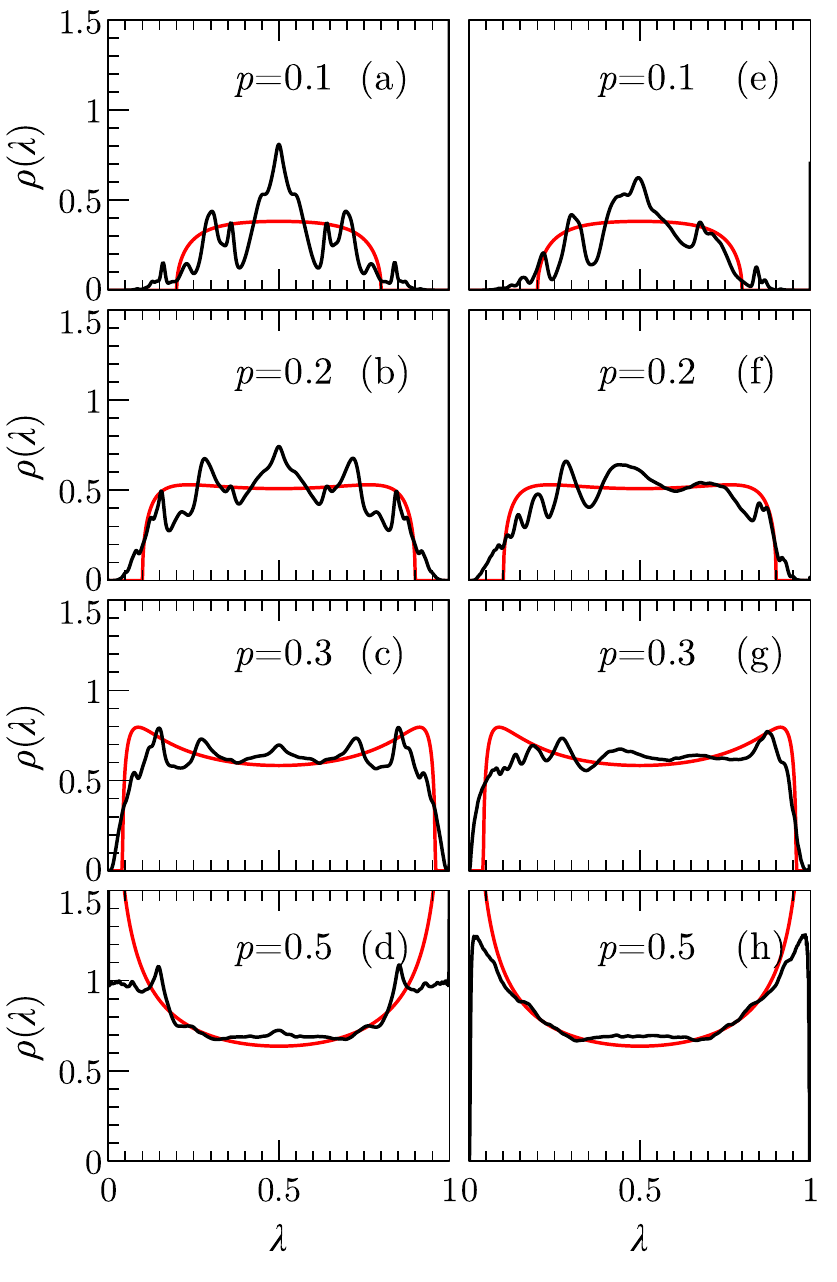}
    \caption{(Color online) Numerically obtained DOS for networks with the topology of (a)-(d) a square lattice and (e)-(h) a diamond lattice with binary distributed bonds at different metal fillings $p$ (black line). The red (gray) line shows the RMT prediction given by Eq. (\ref{eq:RMT_DOS}) with the corresponding $p$ and $z=4$. Numerical calculations are performed for networks of the size $100^2$ and $30^3$ sites, correspondingly, and are averaged over 1000 network configurations using the Kernel polynomial method~\cite{2006_Weisse, 2016_Beltukov}.}
    \label{fig:DOS_Square_RMT}
\end{figure}

\emph{Comparison with numerical results.} First, we consider networks with the topology of a two-dimensional square lattice. This case is the most studied and widely addressed in the literature (see the review \cite{2000_Sarychev} and references therein). The corresponding density of states for a square lattice with different fractions of metallic bonds is shown in Figs.~\ref{fig:DOS_Square_RMT}(a)-\ref{fig:DOS_Square_RMT}(d).

At low filling fractions $p$, the numerically obtained DOS demonstrates the presence of a rich structure with well-resolved resonant peaks. These peaks correspond to resonances of typical clusters which are formed by several metallic bonds embedded into a dielectric lattice --- the so-called {\it lattice animals} \cite{1998_Jonckheere}. For example, the most salient peak at $\lambda=1/2$ corresponds to the dipole resonance of a single metallic bond surrounded by a dielectric environment. In this case, the potential distribution is $\varphi(r) \propto 1/r^2$, which corresponds to a dipole in the two-dimensional electrodynamics \cite{1996_Clerc}. In the dilute case with $p \ll 1$ it is the only remaining peak. Nearby peaks are aligned symmetrically and correspond to resonances of two-bond clusters, and so on. Detailed maps of the resonances of animals on a square lattice can be found in Refs.~\cite{1996_Clerc, 1998_Jonckheere, 2003_Raymond}.

Peaks near the edges of the resonance spectrum are associated with complicated clusters formed by many bonds. The probability of such cluster configurations to occur is low, thus, these peaks are much less pronounced than the central ones. Finally, at the very edges of the spectrum, the amount of resonances is exponentially small, because these resonances are associated with long linear chains of connected metallic bonds which arise with an exponentially small probability \cite{1998_Jonckheere}. Such a behavior is referred to as Lifshitz tails after Lifshitz, who was the first to describe analogous phenomena in the vibrational spectra of binary harmonic alloys \cite{1964_Lifshitz}. As seen from Fig.~\ref{fig:DOS_Square_RMT}, the mentioned peaks are absent in the DOS given by the RMT approach. This originates from our neglect of the correlation between the matrices $M$ and $D$. Indeed, the concept of a cluster loses its meaning in this case. As a result, Lifshitz tails are absent as well. They are replaced by resonance gaps at $0<\lambda<\lambda_{-}(p)$ and $\lambda_{+}(p)<\lambda<1$.

\begin{figure}[t]
    \includegraphics[width=8cm]{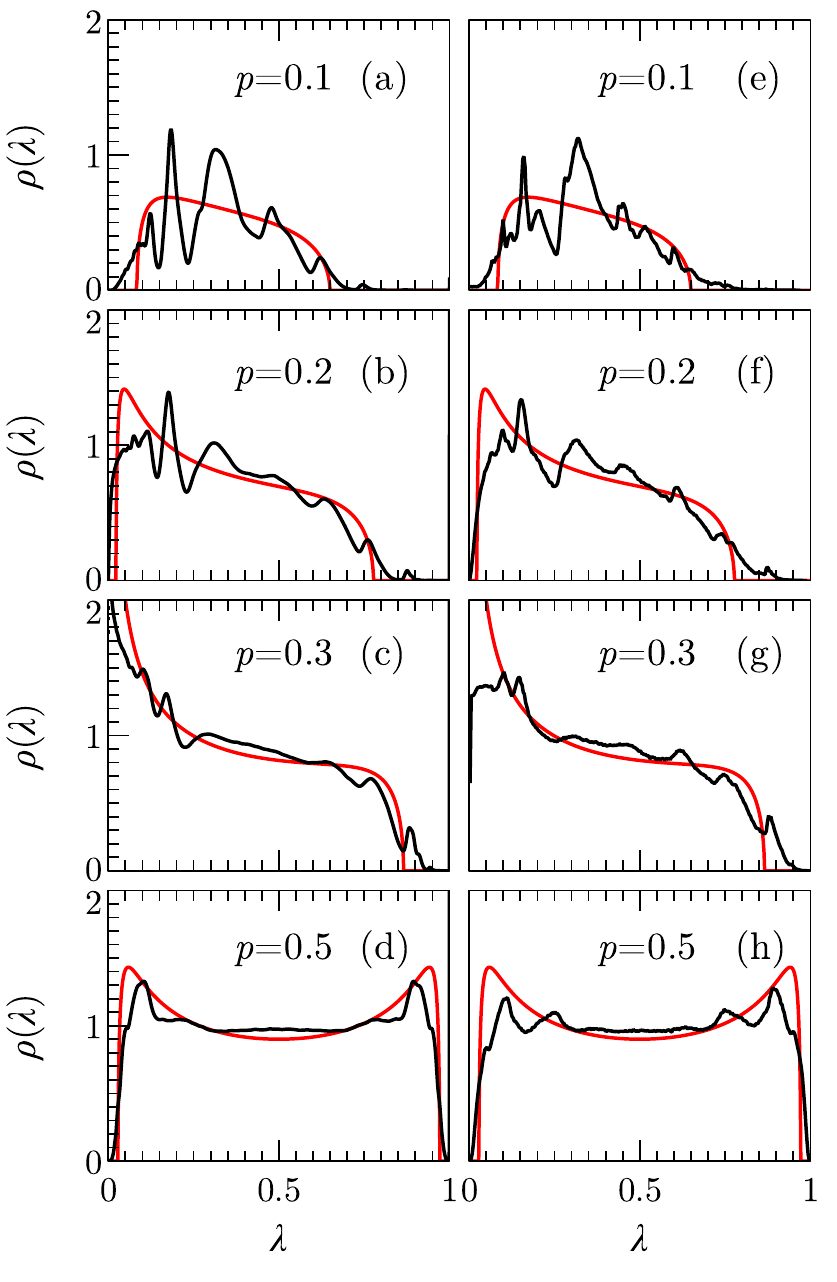}
    \caption{(Color online) Numerically obtained DOS for networks with the topology of (a)-(d) a simple cubic lattice and (e)-(h) a triangular lattice with binary distributed bonds at different metal fillings $p$ (black line). The red (gray) line shows the RMT prediction given by Eq. (\ref{eq:RMT_DOS}) with the corresponding $p$ and $z=6$. Numerical calculations are performed for networks of the size $30^3$ and $100^2$ sites, correspondingly, and are averaged over 1000 network configurations using the Kernel polynomial method~\cite{2006_Weisse, 2016_Beltukov}.}
    \label{fig:DOS_Cubic_RMT}
\end{figure}

At higher fillings $p$, the resonant peaks are less pronounced because of a complication of the geometric structure, which causes typical clusters to be less expressed. Indeed, if $p$ is large enough, then typical clusters usually interact with nearby clusters and are unlikely to be positioned in a large gap filled with dielectric bonds. As a result, the numerical DOS becomes smoother and more similar to the one given by the RMT approach [Figs.~\ref{fig:DOS_Square_RMT}(b) and ~\ref{fig:DOS_Square_RMT}(c)].

As the filling fraction $p$ increases, the bond percolation threshold is reached at some point $p_c$ [Fig.~\ref{fig:DOS_Square_RMT}(d)]. The percolation is a geometric phase transition, which means that at fillings $p>p_c$ an infinite metallic cluster is formed that connects the opposite sides of the system. Thus, an initially insulating system becomes a conducting one in the stationary (${\rm dc}$) regime at $\omega=0$. The percolation threshold can be obtained within the RMT approach as follows. Since resonances correspond to the poles of the conductivity of a system, a nonvanishing dc conductivity corresponds to a nonvanishing DOS at $\omega=0$. Hence, $\lambda_{-}(p_c)=0$ \cite{1992_Bergman}, which gives the RMT estimate of the percolation threshold,
\begin{equation}
    p_c^{RMT}=2/z.
\end{equation}
This result is well known and has a transparent physical interpretation. Indeed, the necessary condition for the existence of a path connecting opposite sides of the network is that at least two of the $z$ bonds which connect a site with its neighbors are metallic. Also,  this estimate gives the exact value $p_c^{\rm sq}=1/2$ for a square lattice.

This result is the only analytically established percolation threshold for a lattice. Its derivation is based upon a special symmetry of the square lattice --- the {\it self-duality}. This peculiar property also causes a mirror symmetry of the DOS under the transform $\lambda \to 1-\lambda$, clearly seen in Figs.~\ref{fig:DOS_Square_RMT}(a)-\ref{fig:DOS_Square_RMT}(d) \cite{1971_Dykhne, 1998_Jonckheere}. The DOS for a diamond lattice is shown for comparison in Figs.~\ref{fig:DOS_Square_RMT}(e)-~\ref{fig:DOS_Square_RMT}(h). This lattice also has the coordination number $z=4$, hence it is described by the same RMT curve. However, it is not self-dual, and, as a result, the corresponding symmetry of $\rho(\lambda)$ is absent. The percolation threshold in the diamond lattice also differs slightly  from that of the square lattice and equals $p_c^{\rm d} = 0.39$.

Next, we compare the results for lattices with the coordination number $z=6$. The DOS for the simple cubic and the triangular lattices at different fillings $p$ is shown in Fig. \ref{fig:DOS_Cubic_RMT}. The whole situation is similar to the previous case with $z=4$. However, there is a difference in frequencies of the dipole resonances of individual metallic bonds, so that $\lambda_{\rm dip}=1/3$. It also corresponds to the frequency of the dipole localized plasmon resonance in a metallic sphere $\omega_p/\sqrt{3}$. This is in agreement with the fact that the dipole potential in the simple cubic lattice decreases with distance as $1/r^3$ \cite{1996_Clerc}. Resonances of other lattice animals differ as well, also due to the different geometries of typical clusters and different probabilities of their occurrence. A map of the resonances of typical clusters in a simple cubic lattice can be found in Ref.~\cite{2000_Albinet}. The percolation thresholds in simple cubic and triangular lattices are $p_c^{\rm sc}=0.25$ and $p_c^{\rm t}=0.35$, correspondingly, which are still close to the RMT prediction $p_c=1/3$.

As was pointed out in Ref.~\cite{1998_Jonckheere}, some of the extensively degenerate eigenvalues of Eq. (\ref{eq:GEP}) with $\lambda=0$ and $\lambda=1$ do not correspond to resonances. A number of these nonphysical eigenvalues is defined by a number of connected clusters formed by dielectric and metallic bonds, correspondingly \cite{2009_Korniss}, and can be easily obtained for any particular implementation of a network.

\emph{Discussion and conclusions.} Let us also point out an interesting interplay between the results given by our RMT approach and by the effective medium approximation (EMA). The latter was introduced by Bruggeman as a self-consistent homogenization scheme for the evaluation of the conductivity of mixtures \cite{1935_Bruggeman, 1992_Bergman} and is widely applied to systems at finite frequencies \cite{1998_Jonckheere, 1973_Kirkpatrick, 1987_Koss, 1993_Zhang, 1995_Zhang}. The main equation of the EMA on a hypercubic lattice reads as \cite{1992_Bergman}
\begin{equation}
    p\frac{\sigma_m - \sigma_{\rm eff}}{\sigma_m + (\frac{z}{2} - 1)\sigma_{\rm eff}} + (1 - p)\frac{\sigma_d - \sigma_{\rm eff}}{\sigma_d + (\frac{z}{2} - 1)\sigma_{\rm eff}}=0,
\end{equation}
where $\sigma_{\rm eff}$ is an effective conductance of the lattice with randomly arranged bonds of conductances $\sigma_m$ and $\sigma_d$. The above equation has an explicit solution which is nonvanishing over the interval $\lambda_{-}^{\rm EMA}<\lambda<\lambda_{+}^{\rm EMA}$, with $\lambda_{\pm}^{\rm EMA}$ given by exactly the same expressions as in the RMT approach, $\lambda_{\pm}^{\rm EMA} = p + \frac{2-4p}{z} \pm \frac{2}{z}\sqrt{2p(1-p)(z-2)}$. Indeed, resonances of the system are poles of its conductance, and thus in a nondissipating system $\rho(\lambda)$ and $\sigma(\lambda)$ should be nonvanishing in the same spectral region. Some correspondence between the random matrix theory and the effective medium description in the case of Gaussian ensembles has been addressed in Refs.~\cite{1999_Biroli, 2002_Semerjian, 2003_Dorogovtsev, 2017_Circuta}.

Predictions of the considered model are in qualitative agreement with the results of recent experiments with lithographic networks \cite{2015_Gaio} and disordered nanocomposite films \cite{2014_Hedayati}. In particular, experimentally measured Purcell enhancement and absorption spectra demonstrate the presence of a broad maximum whose width depends on the metal filling $p$, as well as the presence of an optimal filling which maximizes the absorption band.

To conclude, we have considered a description of resonances in random impedance networks based on the Jacobi ensemble of the random matrix theory. The obtained expressions satisfy all natural symmetries of the considered problem and demonstrate good agreement with the results of numerical simulations, as well as a correspondence with the effective medium approximation. A further development of the obtained description, e.g., a comprehensive study of level spacing statistics \cite{2002_Gu_1, 2007_Lansey} and the properties of eigenvectors, can be of major interest in the area of Anderson localization \cite{1999_Sarychev, 2017_Murphy}.

\emph{Acknowledgments.} We are grateful to V.I. Kozub, D.A. Parshin and D.F. Kornovan for fruitful discussions. This work was financially supported by the Russian Foundation for Basic Research (Project no. 16-32-00359), the ``Dynasty'' Foundation, and Government of Russian Federation (Grant 08-08).


\begin{thebibliography}{99}

\bibitem{Le_Ru_2009} E.C. Le Ru and P. G. Etchegoin, \emph{Principles of Surface Enhanced Raman Spectroscopy and related plasmonic effects} (Elsevier, Amsterdam, 2009).
\bibitem{2001_Breit} M. Breit, V.A. Podolskiy, S. Gresillon, G. von Plessen, J. Feldmann, J.C. Rivoal, P. Gadenne, A.K. Sarychev and V.M. Shalaev, Experimental observation of percolation-enhanced nonlinear light scattering from semicontinuous metal films, Phys. Rev. B {\bf 64}, 125106 (2001).
\bibitem{2015_Carminati} R. Carminati, A. Caze, D. Cao, F. Peragut, V. Krachmalnicoff, R. Pierrat, Y. De Wilde, Electromagnetic density of states in complex plasmonic systems, Surf. Sci. Rep. {\bf 70}, 1 (2015).
\bibitem{2000_Sarychev} A.K. Sarychev and V.M. Shalaev, Electromagnetic field fluctuations and optical nonlinearities in
metal-dielectric composites, Physics Reports {\bf 335}, 275 (2000).
\bibitem{1973_Kirkpatrick} S. Kirkpatrick, Percolation and conduction, Rev. Mod. Phys. {\bf 45}, 574 (1973).
\bibitem{1990_Clerc} J.P. Clerc, G. Giraud, and J.M. Luck, The electrical conductivity of binary disordered systems, percolation clusters, fractals and related models, Adv. Phys. {\bf 39}, 191 (1990).
\bibitem{1992_Bergman} D.J. Bergman, D. Stroud, Physical properties of macroscopically inhomogeneous media, Solid State Physics {\bf 46}, 147 (1992).
\bibitem{1997_Brouers} F. Brouers, S. Blacher, A.N. Lagarkov, A.K. Sarychev, P. Gadenne and V.M. Shalaev, Theory of giant Raman scattering from semicontinuous metal films, Phys. Rev. B {\bf 55}, 13234 (1997).
\bibitem{1998_Brouers} F. Brouers, S. Blacher, A.K. Sarychev, Giant field fluctuations and anomalous light scattering from semicontinuous metal films, Phys. Rev. B {\bf 58}, 15897 (1998).
\bibitem{1998_Jonckheere} Th. Jonckheere, J.M. Luck, Dielectric resonances of binary random networks, J. Phys. A {\bf 31}, 3687 (1998).
\bibitem{2000_Albinet} G. Albinet and L. Raymond, Dielectric resonances in three-dimensional binary disordered media, Eur. Phys. J. B {\bf 13}, 561 (2000).
\bibitem{2012_Murphy} N.B. Murphy and K.M. Golden, The Ising model and critical behavior of transport in binary composite media, J. Math. Phys. {\bf 53}, 063506 (2012).
\bibitem{1987_Koss} R.S. Koss and D. Stroud, Scaling behavior and surface-plasmon modes in metal-insulator composites, Phys. Rev. B {\bf 35}, 9004 (1987).
\bibitem{1993_Zhang} X. Zhang and D. Stroud, Scaling behavior and surface-plasmon resonances in a model three-dimensional metal-insulator composite, Phys. Rev B {\bf 48}, 6658 (1993).
\bibitem{1995_Zhang} X. Zhang and D. Stroud, Optical and electrical properties of thin films, Phys. Rev. B {\bf 52}, 2131 (1995).
\bibitem{1999_Sarychev} A.K. Sarychev, V.A. Shubin and V.M. Shalaev, Anderson localization of surface plasmons and nonlinear optics of metal-dielectric composites, Phys. Rev. B {\bf 60}, 16389 (1999).
\bibitem{2017_Murphy} N.B. Murphy, E. Cherkaev, and K.M. Golden, Anderson Transition for Classical Transport in Composite Materials, Phys. Rev. Lett. {\bf 118}, 036401 (2017).
\bibitem{2002_Gu} Y. Gu, K.W. Yu, and Z.R. Yang, Fluctuations and scaling of inverse participation ratios in random binary resonant composites, Phys. Rev. B. {\bf 66}, 012202 (2002).
\bibitem{1962_Dyson} F. J. Dyson, The threefold way. Algebraic structure of symmetry groups and ensembles in quantum mechanics, J. Math. Phys. (N.Y.) {\bf 3}, 1199 (1962).
\bibitem{Haake} F. Haake, \emph{Quantum Signatures of Chaos} (Springer, Berlin, 2001).
\bibitem{Oxford_Handbook} G. Akemann, J. Baik, and P. Di Francesco, {\it The Oxford Handbook of Random Matrix Theory} (Oxford University Press, Oxford, U.K., 2011).
\bibitem{2000_Mirlin} A.D. Mirlin, Statistics of energy levels and eigenfunctions in disordered systems, Phys. Rep. {\bf 326}, 259 (2000).
\bibitem{2017_Li} H. Li, S. Suwunnarat, R. Fleischmann, H. Schanz, and T. Kottos, Random Matrix Theory Approach to Chaotic Coherent Perfect Absorbers, Phys. Rev. Lett. {\bf 118}, 044101 (2017).
\bibitem{2015_Beltukov} Y.M. Beltukov, Random matrix theory approach to vibrations near the jamming transition, JETP Letters {\bf 101}, 345 (2015).
\bibitem{2008_Evers} F. Evers and A. D. Mirlin, Anderson transitions, Rev. Mod. Phys. {\bf 80}, 1355 (2008).
\bibitem{1977_Webman} I. Webman, J. Jortner, M.H. Cohen, Theory of optical and microwave properties of microscopically inhomogeneous materials, Phys. Rev. B {\bf 15}, 5712 (1977).
\bibitem{LL8} L.D. Landau,  L. P. Pitaevskii and E.M. Lifshitz, \emph{Electrodynamics of continuous media}, 2nd ed. (Butterworth-Heinemann, Oxrofd, U.K., 1984).
\bibitem{1996_Clerc} J.P. Clerc, G. Giraud, J.M. Luck and Th. Robin, Dielectric resonances of lattice animals and other fractal
clusters, J. Phys. A {\bf 29}, 4781 (1996).
\bibitem{1989_Zeng} X.C. Zeng, P.M. Hui and D. Stroud, Numerical study of optical absorption in two-dimensional metal-insulator
and normal-superconductor composites, Phys. Rev. B {\bf 39}, 1063 (1989).
\bibitem{2016_OBP} N.A. Olekhno, Y.M. Beltukov, D.A. Parshin, Spectral properties of plasmon resonances in a random impedance
network model of binary nanocomposites, JETP Lett. {\bf 103}, 577 (2016).
\bibitem{2009_Korniss} R. Huang, G. Korniss and S. K. Nayak, Interplay between structural randomness, composite disorder, and electrical response: Resonances and transient delays in complex impedance networks, Phys. Rev. E {\bf 80}, 045101(R) (2009).
\bibitem{2017_Aouaichia} M. Aouaichia, N. McCullen, C.R. Bowen, D.P. Almond, C. Budd, and R. Bouamrane, Understanding the anomalous frequency responses of composite materials using very large random resistor-capacitor networks, Eur. Phys. J. B {\bf 90}, 39 (2017).
\bibitem{1979_Bergman_1} D.J. Bergman, Dielectric constant of a two-component granular composite: A practical scheme for calculating the pole spectrum, Phys. Rev. B {\bf 19}, 2359 (1979).
\bibitem{1979_Bergman_2} D.J. Bergman, The dielectric constant of a simple cubic array of identical spheres, J. Phys. C: Solid State Phys. {\bf 12}, 4947 (1979).
\bibitem{Horn-book} R.A. Horn, C.R. Johnson, \emph{Matrix Analysis} (Cambridge University Press, Cambridge, U.K., 1990).
\bibitem{Ortega-book} J.M. Ortega, \emph{Matrix Theory: A Second Course} (Springer, Berlin, 1987).
\bibitem{1999_Fyodorov} Y.V. Fyodorov, Spectral properties of random reactance networks and random matrix pencils, J. Phys. A: Math. Gen. {\bf 32}, 7429 (1999).
\bibitem{1999_Fyodorov_1} Y.V. Fyodorov, Fluctuations in random $RL$-$C$ networks: Nonlinear $\sigma$-model description, JETP Letters {\bf 70}, 743 (1999).
\bibitem{2001_Fyodorov} Y.V. Fyodorov, Long-ranged model of random $RL$-$C$ network, Physica E {\bf 9}, 609 (2001).
\bibitem{2003_Staring} J. St\"aring, B. Mehlig, Y.V. Fyodorov, J.M. Luck, On random symmetric matrices with a constraint: the spectral density of random impedance networks, Phys. Rev. E {\bf 67}, 047101 (2003).
\bibitem{Bollobaas} B. Bolloba\'as, {\it Modern Graph Theory} (Springer, Berlin, 1998).
\bibitem{Forrester} P.J. Forrester, {\it Log-Gases and Random Matrices} (Princeton University Press, Princeton, NJ, 2010).
\bibitem{1963_Constantine} A.G. Constantine, Some non-central distribution problems in multivariate analysis, Ann. Math. Stat. {\bf 34}, 1270 (1963).
\bibitem{2013_Erdos} L. Erdos, B. Farell, Local eigenvalue density for general MANOVA matrices, J. Stat. Phys. {\bf 152}, 1003 (2013).
\bibitem{2003_Raymond}  L. Raymond, J.M. Laugier, S. Schafer and G. Albinet, Dielectric resonances in disordered media, Eur. Phys. J. B {\bf 31}, 355 (2003).
\bibitem{1964_Lifshitz} I.M. Lifshitz, The energy spectrum of disordered systems, Adv. Phys. {\bf 13}, 483 (1964).
\bibitem{2006_Weisse} A. Wei{\ss}e, G. Wellein, A. Alvermann, H. Fehske, The kernel polynomial method, Rev. Mod. Phys. {\bf 78}, 275 (2006).
\bibitem{2016_Beltukov} Y. M. Beltukov, C. Fusco, D. A. Parshin, and A. Tanguy, Boson peak and Ioffe-Regel criterion in amorphous siliconlike materials: The effect of bond directionality, Phys. Rev. E {\bf 93}, 023006 (2016).
\bibitem{1971_Dykhne} A.M. Dykhne, Conductivity of a two-dimensional two-phase system, Sov. Phys. JETP {\bf 32}, 63 (1971).
\bibitem{1935_Bruggeman} D.A.G. Bruggeman, Berechnung verschiedener physikalischer Konstanten von heterogenen Substanzen. I. Dielektrizit{\"a}tskonstanten und Leitf{\"a}higkeiten der Mischk{\"o}rper aus isotropen Substanzen, Ann. Phys. (Leipzig, Ger.) {\bf 24}, 636 (1935).
\bibitem{1999_Biroli} G. Biroli and R. Monasson, A single defect approximation for localized states on random lattices, J. Phys. A: Math. Gen. {\bf 32} L255 (1999).
\bibitem{2002_Semerjian} G. Semerjian and F. Cugliandolo, Sparse random matrices: the eigenvalue spectrum revisited, J. P
hys. A: Math. Gen. {\bf 35}, 4837 (2002).
\bibitem{2003_Dorogovtsev} S.N. Dorogovtsev, A.V. Goltsev, J.F.F. Mendes, and A.N. Samukhin, Spectra of complex networks, Phys. Rev. E {\bf 68}, 046109 (2003).
\bibitem{2017_Circuta} G.M. Cicuta, J. Krausser, R. Milkus, and A. Zaccone, Unifying model for random matrix theory in arbitrary space dimensions, Phys. Rev. E {\bf 97}, 032113 (2018).
\bibitem{2015_Gaio} M. Gaio, M. Castro-Lopez, J. Renger, N. van Hulst and R. Sapienza, Percolating plasmonic networks for light emission control, Faraday Discuss. {\bf 178}, 237 (2015).
\bibitem{2014_Hedayati} M.K. Hedayati, F. Faupel, and M. Elbahri, Review of plasmonic nanocomposite metamaterial absorber, Materials {\bf 7}, 1221 (2014).
\bibitem{2002_Gu_1} Y. Gu, K.W. Yu and Z.R. Yang, Statistics of level spacing of geometric resonances in random binary composites, Phys. Rev. E {\bf 65}, 046129 (2002).
\bibitem{2007_Lansey} E. Lansey, A. Lapin, F. Zypman, Level statistics in disordered linear networks, Physica A {\bf 386}, 655 (2007).

\end{thebibliography}
\end{document}